\documentclass[twocolumn]{aastex61}%
\usepackage{amssymb,amsmath,amsfonts}
\usepackage{graphicx,color}
\usepackage[squaren]{SIunits}
\usepackage[english]{babel}
\usepackage{tabularx}
\newcommand{\dif}{\mathrm{d}}%
\newcommand{\tdif}[2]{\frac{\dif#1}{\dif#2}}%
\newcommand{\abs}[1]{\lvert#1\rvert}%
\makeatletter
\def\frontmatter@preabstractspace{48pt}
\makeatother

\shorttitle{Anisotropy in the arrival directions of UHECR}
\shortauthors{Wittkowski and Kampert}

\begin{document}
\title{On the anisotropy in the arrival directions of ultra-high-energy cosmic rays}

\correspondingauthor{David Wittkowski}
\email{david.wittkowski@uni-wuppertal.de}

\author{David Wittkowski}
\affiliation{Department of Physics, Bergische Universit\"at Wuppertal, Gau\ss{}stra\ss{}e 20, D-42097 Wuppertal, Germany}

\author{Karl-Heinz Kampert}
\affiliation{Department of Physics, Bergische Universit\"at Wuppertal, Gau\ss{}stra\ss{}e 20, D-42097 Wuppertal, Germany}

\begin{abstract}
We present results of elaborate four-dimensional simulations of the propagation of ultra-high-energy cosmic rays (UHECR), which are based on a  realistic astrophysical scenario. The distribution of the arrival directions of the UHECR is found to have a pronounced dipolar anisotropy and rather weak higher-order contributions to the angular power spectrum. This finding agrees well with the recent observation of a dipolar anisotropy for UHECR with arrival energies above $8\,\mathrm{EeV}$ by the Pierre Auger Observatory and constitutes an important prediction for other energy ranges and higher-order angular contributions for which sufficient experimental data are not yet available. Since our astrophysical scenario enables simulations that are completely consistent with the available data, this scenario will be a very useful basis for related future studies. 
\end{abstract}


\section{Introduction}
Although ultra-high-energy cosmic rays (UHECR), i.e., cosmic particles with energies $\geq 1 \,\mathrm{EeV}$, have been investigated for more than half a century \citep{Linsley1963,NaganoW2000}, most of the main questions regarding UHECR are still unanswered \citep[see, e.g.,][]{KoteraO2011}. For example, it is not yet known from which sources UHECR originate, what is the chemical composition of the particles at their sources, and how the particles are accelerated \citep{Sigl2001,KachelriessS2006}.
A way to investigate these issues is to make assumptions about the origin, source composition, etc.\ of UHECR, to simulate their
propagation to Earth under these assumptions, and to compare the simulation results with observational data.
Typical observables for comparing the results of simulations and experiments are the energy spectrum, mass spectrum \citep{AabPAOCfos2017}, and distribution of arrival directions of UHECR reaching the Earth.
From these observables, the last one allows the most direct conclusions about the locations of the sources.

In recent years, there have been strong efforts to study the directional distribution of UHECR arriving at Earth and observational hints for an anisotropy in the arrival directions have been reported \citep{AbbasiTAIois2014,AlSamaraiPAO2015, AabPAO2015,AabPAOMras2017}.
However, a statistically significant (significance level $s>5\sigma$) detection of an UHECR anisotropy was not possible until very recently.
The Pierre Auger Collaboration now reported the discovery of a significant dipolar anisotropy ($s=5.2\sigma$) for cosmic particles arriving with energies $E>8\,\mathrm{EeV}$ \citep{AabPAOOoal2017}.
This experimental work represents important progress towards the identification of the sources of UHECR. Still, it has some observational limitations.
First, it focuses on the existence of a nonzero dipole moment in the orientational distribution of the arrival directions, as the statistics of the experimental data does not allow it to significantly prove higher-order multipole moments.
Second, for similar reasons no higher arrival-energy ranges than $E>8\,\mathrm{EeV}$ are taken into account.
The third limitation arises from the observation of UHECR from only a part of the sky \citep{Sommers2001}. By combining the data of the Pierre Auger Observatory with data from the Telescope Array, it has been possible to reach a full sky coverage for energies $E>10\,\mathrm{EeV}$ \citep{AabPAO2014,DelignyPAOTA2015}, but the data for this energy range did not allow one to find a significant ($s>5\sigma$) anisotropy in the arrival directions and up to now there are no corresponding combined data for lower energies.

The goal of this letter is to study the anisotropy in the arrival directions of UHECR comprehensively without these observational limitations.
For this purpose, we simulated the propagation of UHECR from their assumed sources to Earth, taking into account deflections of the trajectories of charged particles in extragalactic and galactic magnetic fields, all relevant interactions with the photon background, as well as cosmological effects such as the redshift evolution of the photon background and the adiabatic expansion of the universe.
These four-dimensional simulations are limited neither to a specific energy range nor to the consideration of particular multipole moments of the distribution of the arrival directions of UHECR. To study the anisotropy in the arrival directions, we consider the associated angular power spectrum up to order $64$ and its dependence on the arrival energies of the particles. Furthermore, the simulations correspond to a full sky coverage.

With these features, our work can provide guidance for future experimental studies. In contrast to earlier simulation work \citep{Taylor2014,TinyakovU2015,EichmannBTM2017}, our results are in excellent agreement with the energy spectrum, mass spectrum, as well as anisotropy of the current UHECR data collected by the Pierre Auger Observatory.
In addition, the results lead to interesting predictions for energy ranges and multipole moments that are not yet accessible by the existing observatories.

\section{\label{sec:Methods}Methods}
Our four-dimensional simulations of the propagation of UHECR to Earth were carried out using the Monte-Carlo code CRPropa 3 \citep{BatistaDEKKMSVWW2016}. These simulations take into account all three spatial degrees of freedom as well as the cosmological time-evolution of the universe. Regarding the sources, we assumed that they all have the same properties and that they are discrete objects, whose spatial distribution follows the local mass distribution of the universe. We chose their positions randomly such that the local large-scale mass structure resembles the model of \citet{DolagGST2005}. 
The minimal source distance from the observer was $10 \,\mathrm{Mpc}$ to avoid effects of nearby sources and the maximal redshift was $z\approx 1.3$, which is equivalent to a maximal comoving distance of $4\,\mathrm{Gpc}$. For the source density we chose $\rho \approx 10^{-4}\,\mathrm{Mpc^{-3}}$, which is in accordance with known density bounds \citep{AbreuPAO2013dens}.

In our simulations, the sources emitted $^1\mathrm{H}$, $^4\mathrm{He}$, $^{14}\mathrm{N}$, $^{28}\mathrm{Si}$, and $^{56}\mathrm{Fe}$ isotropically with an energy spectrum 
\begin{equation}
J_{0}(E_{0})=\tdif{N_{0}}{E_{0}} \propto \sum_{\alpha} f_{\alpha} E_{0}^{-\gamma}
\begin{cases}%
1\,, & \text{if}\; \frac{E_{0}}{Z_{\alpha}}<R_{\mathrm{cut}}\,, \\
e^{1-\frac{E_{0}}{Z_{\alpha} R_{\mathrm{cut}}}}\,, & \text{if}\; \frac{E_{0}}{Z_{\alpha}}\geq R_{\mathrm{cut}} \,,
\end{cases}%
\end{equation}
where $\dif N_{0}(E_{0})$ is the number of particles emitted with an energy in the interval from $E_{0}$ to $E_{0}+\dif E_{0}$.
Here, $f_{\alpha}$ is the fraction of particles of element $\alpha\in\{\mathrm{H}, \mathrm{He}, \mathrm{N}, \mathrm{Si},\mathrm{Fe}\}$ among all emitted particles with the normalization $\sum_{\alpha} f_{\alpha} = 1$, $\gamma$ is the so-called spectral index, $Z_{\alpha}$ is the atomic number of element $\alpha$, and $R_{\mathrm{cut}}$ is a cut-off rigidity above which the particle flow at the sources is exponentially suppressed. 
We chose the source parameters $f_{\alpha}$, $\gamma$, and $R_{\mathrm{cut}}$ as $f_{\mathrm{H}}=3.0\%$, $f_{\mathrm{He}}=2.1\%$, $f_{\mathrm{N}}=73.5\%$, $f_{\mathrm{Si}}=21.0\%$, $f_{\mathrm{Fe}}=0.4\%$, $\gamma=1.61$, and $R_{\mathrm{cut}}=10^{18.88} \mathrm{eV}$, since for these parameter values the energy spectrum and mass spectrum of the simulated UHECR arriving at the observer are in optimal agreement with the corresponding data from the Pierre Auger Observatory \citep{WittkowskiDPAO2017}.
To obtain good statistics, we emitted more than $10^9$ particles at the sources.
These particles were only charged nuclei and not, e.g., photons or neutrinos, since experiments have shown that UHECR mainly consist of charged nuclei \citep{AbrahamPAO2009,AbreuPAOII2013,AbuTAC2013,AloisioBDGPS2015}. 

When simulating the propagation of UHECR, their trajectories are influenced by interactions with the extragalactic photon background, by deflections in extragalactic and galactic magnetic fields, as well as by cosmological effects like the redshift evolution of the photon background and the adiabatic expansion of the universe.  
For the extragalactic background light, we used the model of \citet{Gilmore2012} (the so-called ``fiducial'' model) as well as the photodisintegration cross sections from the TALYS code\footnote{\url{http://www.talys.eu/documentation/}} \citep{KoningHD2005,KoningR2012}. 
Moreover, we applied the same extragalactic magnetic field model as in \citet{BatistaDEKKMSVWW2016} together with reflective boundary conditions  \citep{Haghighat2015}. 
The particles were propagated through the extragalactic magnetic field until they hit a sphere of radius $1\,\mathrm{Mpc}$ that was centered at Earth and captured all particles arriving with redshift $-0.025 < z < 0.025$. Next, the effect of the galactic magnetic field on the particles was calculated using the JF 2012 model of Jansson and Farrar \citep{JanssonFAnmo2012,JanssonFTgmf2012,BatistaDEKKMSVWW2016} for the galactic magnetic field. We checked that using a smaller sphere does not change the simulation results qualitatively and used the sphere of radius $1\,\mathrm{Mpc}$ for better statistics.

To study the distribution of the arrival directions, they were binned into a HEALPix grid\footnote{\url{http://healpix.jpl.nasa.gov/}} \citep{GorskiHBWHRB2005healpix} of $49152$ cells of the same solid angle. This resulted in a coarse-grained distribution $\mathcal{N}(E,\boldsymbol{\hat{n}})$ of the number of detected particles $\mathcal{N}$ as a function of their arrival energy $E$ and arrival direction $\boldsymbol{\hat{n}}$, where the unit vector $\boldsymbol{\hat{n}}$ corresponds to the sign-inversed and normalized momentum vector of an arriving particle. Through the choice of $49152$ cells, the angular resolution of the coarse-grained distribution was similar to the angular resolution of the Pierre Auger Observatory \citep{BonifaziPAO2009} in the relevant energy range. 
We expanded the rescaled particle number distribution $(\mathcal{N}(E,\boldsymbol{\hat{n}})-\langle\mathcal{N}\rangle(E))/\langle\mathcal{N}\rangle(E)$ with $\langle\,\cdot\,\rangle$ denoting an angular average, i.e., the relative fluctuations in the particle number, into spherical harmonics $Y_l^m(\boldsymbol{\hat{n}})$:
\begin{equation}
\frac{\mathcal{N}(E,\boldsymbol{\hat{n}}) - \langle\mathcal{N}\rangle(E)}{\langle\mathcal{N}\rangle(E)} = \sum_{l=0}^{l_{\mathrm{max}}} \sum^{l}_{m=-l} a_{lm}(E)Y_l^m(\boldsymbol{\hat{n}}) \;.
\end{equation}
Here, $l_{\mathrm{max}}$ is the maximal order of the expansion we were interested in and $a_{lm}(E)$ are the energy-dependent expansion coefficients.
The angular power spectrum corresponding to the distribution of the arrival directions of the simulated UHECR is then given by 
\begin{equation}
C_l(E) = \frac{1}{2l+1} \sum_{m=-l}^{l} \abs{a_{lm}(E)}^2 
\end{equation}
with $l\in\{0,\dotsc,l_{\mathrm{max}}\}$, where $C_0(E)=0$ due to the rescaling of $\mathcal{N}(E,\boldsymbol{\hat{n}})$. 
Note that the coefficients $C_l(E)$ are energy-dependent and rotationally invariant. For $l\geq 1$, they describe the angular distribution of the arrival directions on solid angle scales $2\pi/l\,\mathrm{sr}$. The angular power spectrum is therefore a useful quantity to study the distribution of the arrival directions and to find possible anisotropies. 

To see which coefficients $C_l(E)$ can be measured in the near future with statistical significance $s>5\sigma$, we determined the upper $5\sigma$ confidence bounds for isotropy. For this purpose, we estimated that in a few years about $50000$,  $34988$, and $18288$ UHECR events with energies greater than $8\,\mathrm{EeV}$, $10\,\mathrm{EeV}$, and $15\,\mathrm{EeV}$, respectively, will have been detected by UHECR observatories. 
Furthermore, for each of these three energy intervals we generated $10^{7}$ data sets of $50000$, $34988$, and $18288$ UHECR events, respectively, with random arrival directions that are uniformly distributed on the unit sphere.
From these data sets we then determined the mean values and standard deviations $\sigma$ of the coefficients $C_l$, which allowed us to calculate the upper $5\sigma$ confidence bound for isotropy.

\section{\label{sec:Results}Results}
\begin{figure}[ht]
\begin{center}%
\includegraphics[width=\linewidth]{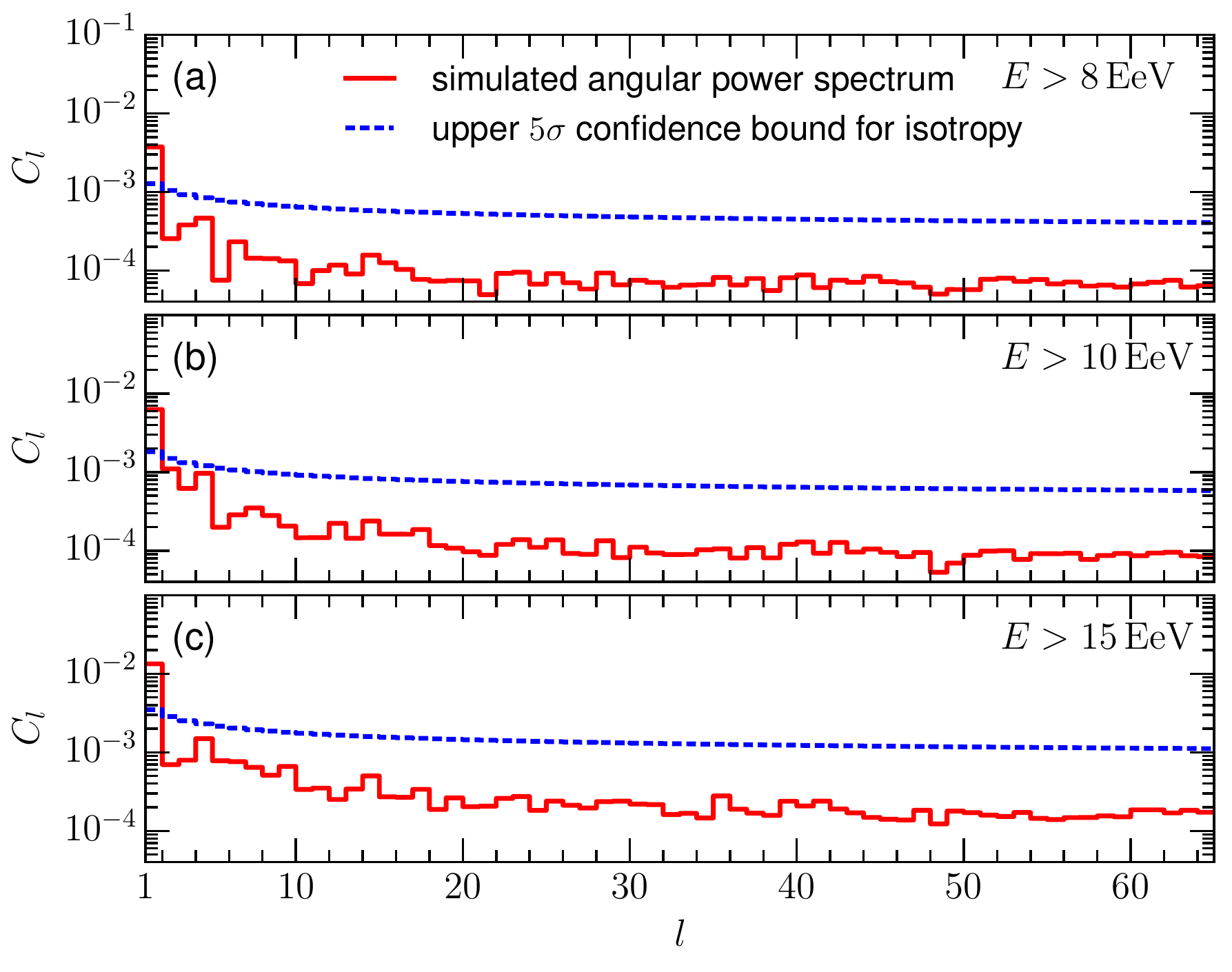}%
\caption{\label{f1}Angular power spectrum (solid red curves) for the arrival directions of the simulated UHECR reaching the observer with energies (a) $E>8\,\mathrm{EeV}$, (b) $E>10\,\mathrm{EeV}$, and (c) $E>15\,\mathrm{EeV}$ as well as the corresponding upper $5\sigma$ confidence bounds for isotropy (dashed blue curves). 
For all energy intervals there is a significant dipolar anisotropy (see the values of $C_1(E)$), whereas the higher-order $C_l(E)$ are compatible with isotropy.}%
\end{center}%
\end{figure}

Figure \ref{f1} shows the angular power spectrum of the arrival directions of the simulated UHECR up to order $l_{\mathrm{max}}=64$ for particles reaching the Earth with energies $E>8\,\mathrm{EeV}$, $E>10\,\mathrm{EeV}$, or $E>15\,\mathrm{EeV}$.
Remarkably, for all three energy ranges our results show a statistically significant ($s>5\sigma$) dipolar anisotropy, whereas on smaller solid angle scales the distribution of the arrival directions is always compatible with an isotropic directional distribution.   
The significant dipolar anisotropy with no higher-order anisotropies for arrival energies $E>8\,\mathrm{EeV}$ (see Fig.\ \ref{f1}a) is in excellent agreement with the latest data of the Pierre Auger Observatory. 
In a recent study, the Pierre Auger Collaboration reported a significant ($s>5\sigma$) dipolar anisotropy \citep{AabPAOOoal2017}, whereas an earlier study found isotropy for the higher-order multipole moments \citep{AabPAOMras2017}. 
Similarly, our results for arrival energies $E>10\,\mathrm{EeV}$ (see Fig.\ \ref{f1}b) are well in line with \citet{DelignyPAOTA2015}, which mentions indications of a dipolar anisotropy for the same energy range. This finding, however, is not statistically significant with reference to a significance level of $5\sigma$, but based on our simulation results we can expect that the dipolar anisotropy in the experimental data will become significant as soon as enough data are available. \citet{DelignyPAOTA2015} also considers higher-order contributions to the angular power spectrum up to order $l=20$ and finds that they are compatible with isotropy. This is again consistent with our simulations and we predict that one will find no significant anisotropy also for even larger $l$ when they are addressed in experiments in the near future.    
Our results for arrival energies $E>15\,\mathrm{EeV}$ (see Fig.\ \ref{f1}c) are predictions that we expect to get confirmed by future experimental studies. We are not aware of any relevant studies considering the angular power spectrum of the arrival directions of UHECR in this energy range. 

For $E>8\,\mathrm{EeV}$, $E>10\,\mathrm{EeV}$, and $E>15\,\mathrm{EeV}$ the values of $C_1(E)$ are $3.74 \cdot 10^{-3}$, $6.28 \cdot 10^{-3}$, and $1.34 \cdot 10^{-2}$, respectively. The corresponding dipole amplitudes $\frac{3}{2\sqrt{\pi}} \sqrt{C_1(E)}$ are approximately $5.2 \cdot 10^{-2}$, $6.7\cdot 10^{-2}$, and $9.8\cdot 10^{-2}$, respectively.
Notably, the dipole amplitudes for $E>8\,\mathrm{EeV}$ and $E>10\,\mathrm{EeV}$ are close to the experimentally determined ones \citep{AabPAOOoal2017, DelignyPAOTA2015}. 
The values of $C_4(E)$ are noticeable for all investigated energy intervals, but will not be measurable with significance $s>5\sigma$ in the next few years with the current UHECR observatories.  

An important feature of our simulation results is the fact that they are in very good agreement with the energy spectrum and mass spectrum as well as with the angular power spectrum of the corresponding experimental data that are currently available.  
Earlier studies have usually focused only on either the energy spectrum and mass spectrum or on anisotropies in the arrival directions of UHECR, but did not present a consistent explanation for all three of these observables \citep{KoteraO2011,Taylor2014,TinyakovU2015,TakamiIY2012}. The  few existing exceptions had difficulties to simultaneously reproduce the experimental results for all observables.  
An example is the recently published work \citet{EichmannBTM2017} that obtained a too strong anisotropy in the arrival directions.  
In contrast to these other studies, our work is apparently based on a more realistic astrophysical scenario that enables simulations that are completely consistent with the available experimental UHECR data and allow to make predictions for energy ranges and contributions to the angular power spectrum for which sufficient experimental data are not yet available.

\section{\label{sec:Conclusions}Conclusions}
We have simulated the propagation of UHECR from their assumed sources to Earth and investigated the distribution of their arrival directions.
For this purpose, we carried out elaborate four-dimensional simulations that are based on a realistic astrophysical scenario, which assumes that the distribution of the sources follows the local large-scale mass structure described by the model of \citet{DolagGST2005}. 
The results of our simulations are in excellent agreement with the available UHECR data collected by the Pierre Auger Observatory. Regarding the distribution of the arrival directions, we found a remarkable dipolar anisotropy and rather weak higher-order multipole moments. The dipolar anisotropy is energy-dependent, but clearly pronounced for all arrival-energy ranges we considered ($E>8\,\mathrm{EeV}$, $E>10\,\mathrm{EeV}$, and $E>15\,\mathrm{EeV}$). 

These findings agree well with the recent observation of a significant dipolar anisotropy \citep{AabPAOOoal2017}, and no significant departure from  isotropy for the higher-order multipole moments \citep{AabPAOMras2017}, for arrival energies $E>8\,\mathrm{EeV}$. They are also well in line with indications of a dipolar anisotropy, and higher-order contributions ($2\leq l \leq 20$) to the angular power spectrum that are compatible with isotropy, reported for $E>10\,\mathrm{EeV}$ in a combined study of the Pierre Auger Collaboration and the Telescope Array Collaboration \citep{DelignyPAOTA2015}. For higher energies and higher-order multipole moments there are at present no experimental findings corresponding to our results, which therefore constitute important predictions that we expect to get confirmed as soon as sufficient experimental data are available. 

The excellent agreement of our simulations and related experimental data shows that the astrophysical scenario underlying our simulations is realistic and that it consistently describes the properties of the sources of UHECR, their emission at the sources, and the propagation to Earth. This gives important hints on the still unknown real sources of UHECR and their properties.  
Furthermore, with its outstanding features this astrophysical scenario is a very useful basis for future simulation studies.   
One could use simulations based on this scenario, e.g., to predict the flux of photons that originate from interactions of UHECR with the extragalactic background light. These photons are interesting since they provide additional information about the sources of UHECR, but up to now it was not yet possible to detect photons with the particularly attractive energies $E>1\,\mathrm{EeV}$. Therefore, predictions for the flux of such photons would be very useful. These predictions could be compared with the upper photon flux limits determined by the Pierre Auger Observatory \citep{AabPAOSfpw2017} and would help to design future gamma-ray detectors \citep{Knodlseder2016,Cyranoski2017}.

\begin{acknowledgments}
\section*{Acknowledgments}
The simulations were carried out on the pleiades cluster at the University of Wuppertal, which was supported by the Deutsche Forschungsgemeinschaft (DFG). Further financial support by the BMBF Verbundforschung Astroteilchenphysik is acknowledged.
\end{acknowledgments}

\bibliographystyle{aasjournal}
\bibliography{refs}

\begin{thebibliography}{}
\expandafter\ifx\csname natexlab\endcsname\relax\def\natexlab#1{#1}\fi
\providecommand{\url}[1]{\href{#1}{#1}}

\bibitem[{Aab {et~al.}(2014)}]{AabPAO2014}
Aab, A., {et~al.} 2014, Astrophysical Journal, 794, 172

\bibitem[{Aab {et~al.}(2015)}]{AabPAO2015}
---. 2015, Astrophysical Journal, 802, 111

\bibitem[{Aab {et~al.}(2017{\natexlab{a}})}]{AabPAOCfos2017}
---. 2017{\natexlab{a}}, Journal of Cosmology and Astroparticle Physics, 2017,
  038

\bibitem[{Aab {et~al.}(2017{\natexlab{b}})}]{AabPAOMras2017}
---. 2017{\natexlab{b}}, Journal of Cosmology and Astroparticle Physics, 2017,
  026

\bibitem[{Aab {et~al.}(2017{\natexlab{c}})}]{AabPAOOoal2017}
---. 2017{\natexlab{c}}, Science, 357, 1266

\bibitem[{Aab {et~al.}(2017{\natexlab{d}})}]{AabPAOSfpw2017}
---. 2017{\natexlab{d}}, Journal of Cosmology and Astroparticle Physics, 2017,
  009

\bibitem[{Abbasi {et~al.}(2014)}]{AbbasiTAIois2014}
Abbasi, R.~U., {et~al.} 2014, Astrophysical Journal Letters, 790, L21

\bibitem[{Abraham {et~al.}(2009)}]{AbrahamPAO2009}
Abraham, J., {et~al.} 2009, Astroparticle Physics, 31, 399

\bibitem[{Abreu {et~al.}(2013{\natexlab{a}})}]{AbreuPAO2013dens}
Abreu, P., {et~al.} 2013{\natexlab{a}}, Journal of Cosmology and Astroparticle
  Physics, 2013, 009

\bibitem[{Abreu {et~al.}(2013{\natexlab{b}})}]{AbreuPAOII2013}
---. 2013{\natexlab{b}}, Advances in High Energy Physics, 2013, 18

\bibitem[{Abu-Zayyad {et~al.}(2013)}]{AbuTAC2013}
Abu-Zayyad, T., {et~al.} 2013, Physical Review D, 88, 112005

\bibitem[{Aloisio {et~al.}(2015)Aloisio, Boncioli, {di Matteo}, Grillo,
  Petrera, \& Salamida}]{AloisioBDGPS2015}
Aloisio, R., Boncioli, D., {di Matteo}, A., {et~al.} 2015, Journal of Cosmology
  and Astroparticle Physics, 2015, 006

\bibitem[{Batista {et~al.}(2016)Batista, Dundovic, Erdmann, Kampert, Kuempel,
  M{\"u}ller, Sigl, van Vliet, Walz, \& Winchen}]{BatistaDEKKMSVWW2016}
Batista, R.~A., Dundovic, A., Erdmann, M., {et~al.} 2016, Journal of Cosmology
  and Astroparticle Physics, 2016, 038

\bibitem[{Bonifazi {et~al.}(2009)}]{BonifaziPAO2009}
Bonifazi, C., {et~al.} 2009, Nuclear Physics B-Proceedings Supplements, 190, 20

\bibitem[{Cyranoski(2017)}]{Cyranoski2017}
Cyranoski, D. 2017, Nature, 543, 300

\bibitem[{{D. Wittkowski for the {P}ierre {A}uger
  {C}ollaboration}(2017)}]{WittkowskiDPAO2017}
{D. Wittkowski for the {P}ierre {A}uger {C}ollaboration}. 2017, in Proceedings
  of the 35th {I}nternational {C}osmic {R}ay {C}onference (ICRC 2017),
  Proceedings of Science (Trieste: SISSA), 563, arXiv:1708.06592

\bibitem[{Dolag {et~al.}(2005)Dolag, Grasso, Springel, \&
  Tkachev}]{DolagGST2005}
Dolag, K., Grasso, D., Springel, V., \& Tkachev, I. 2005, Journal of Cosmology
  and Astroparticle Physics, 1, 9

\bibitem[{Eichmann {et~al.}(2017)Eichmann, Becker~Tjus, \&
  Merten}]{EichmannBTM2017}
Eichmann, B., Becker~Tjus, J., \& Merten, L. 2017, preprint, arXiv:1701.06792

\bibitem[{Gilmore {et~al.}(2012)Gilmore, Somerville, Primack, \&
  Dom{\'\i}nguez}]{Gilmore2012}
Gilmore, R.~C., Somerville, R.~S., Primack, J.~R., \& Dom{\'\i}nguez, A. 2012,
  Monthly Notices of the Royal Astronomical Society, 422, 3189

\bibitem[{Gorski {et~al.}(2005)Gorski, Hivon, Banday, Wandelt, Hansen,
  Reinecke, \& Bartelmann}]{GorskiHBWHRB2005healpix}
Gorski, K.~M., Hivon, E., Banday, A.~J., {et~al.} 2005, Astrophysical Journal,
  622, 759

\bibitem[{Haghighat(2016)}]{Haghighat2015}
Haghighat, A. 2016, Monte Carlo Methods for Particle Transport, 1st edn.,
  Vol.~1 (Boca Raton: CRC Press)

\bibitem[{{I. Al Samarai for the {P}ierre {A}uger
  {C}ollaboration}(2015)}]{AlSamaraiPAO2015}
{I. Al Samarai for the {P}ierre {A}uger {C}ollaboration}. 2015, in Proceedings
  of the 34th {I}nternational {C}osmic {R}ay {C}onference (ICRC 2015),
  Proceedings of Science (Trieste: SISSA), 372

\bibitem[{Jansson \& Farrar(2012{\natexlab{a}})}]{JanssonFAnmo2012}
Jansson, R., \& Farrar, G.~R. 2012{\natexlab{a}}, Astrophysical Journal, 757,
  14

\bibitem[{Jansson \& Farrar(2012{\natexlab{b}})}]{JanssonFTgmf2012}
---. 2012{\natexlab{b}}, Astrophysical Journal Letters, 761, L11

\bibitem[{Kachelriess \& Serpico(2006)}]{KachelriessS2006}
Kachelriess, M., \& Serpico, P.~D. 2006, Physics Letters B, 640, 225

\bibitem[{Kn{\"o}dlseder(2016)}]{Knodlseder2016}
Kn{\"o}dlseder, J. 2016, Comptes Rendus Physique, 17, 663

\bibitem[{Koning {et~al.}(2005)Koning, Hilaire, \& Duijvestijn}]{KoningHD2005}
Koning, A.~J., Hilaire, S., \& Duijvestijn, M.~C. 2005, in AIP Conference
  Proceedings, Vol. 769, 1154--1159

\bibitem[{Koning \& Rochman(2012)}]{KoningR2012}
Koning, A.~J., \& Rochman, D. 2012, Nuclear Data Sheets, 113, 2841

\bibitem[{Kotera \& Olinto(2011)}]{KoteraO2011}
Kotera, K., \& Olinto, A.~V. 2011, Annual Review of Astronomy and Astrophysics,
  49, 119

\bibitem[{Linsley(1963)}]{Linsley1963}
Linsley, J. 1963, Physical Review Letters, 10, 146

\bibitem[{Nagano \& Watson(2000)}]{NaganoW2000}
Nagano, M., \& Watson, A.~A. 2000, Reviews of Modern Physics, 72, 689

\bibitem[{{O. Deligny for the {P}ierre {A}uger {C}ollaboration and {T}elescope
  {A}rray {C}ollaboration}(2015)}]{DelignyPAOTA2015}
{O. Deligny for the {P}ierre {A}uger {C}ollaboration and {T}elescope {A}rray
  {C}ollaboration}. 2015, in Proceedings of the 34th International Cosmic Ray
  Conference (ICRC 2015), Proceedings of Science (Trieste: SISSA), 395

\bibitem[{Sigl(2001)}]{Sigl2001}
Sigl, G. 2001, Science, 291, 73

\bibitem[{Sommers(2001)}]{Sommers2001}
Sommers, P. 2001, Astroparticle Physics, 14, 271

\bibitem[{Takami {et~al.}(2012)Takami, Inoue, \& Yamamoto}]{TakamiIY2012}
Takami, H., Inoue, S., \& Yamamoto, T. 2012, Astroparticle Physics, 35, 767

\bibitem[{Taylor(2014)}]{Taylor2014}
Taylor, A.~M. 2014, Astroparticle Physics, 54, 48

\bibitem[{Tinyakov \& Urban(2015)}]{TinyakovU2015}
Tinyakov, P.~G., \& Urban, F.~R. 2015, Journal of Experimental and Theoretical
  Physics, 120, 533

\end{thebibliography}
\end{document}